\documentclass[a4paper]{scrartcl}
\usepackage[utf8]{inputenc}
\usepackage[margin=25mm]{geometry}
\usepackage{lmodern}  
\usepackage{verbatim}
\usepackage{amsmath,amssymb,array,booktabs}
\usepackage{bbm,slashed,graphicx}
\usepackage{url,hyperref}
\providecommand{\keywords}[1]
{\small	
\textbf{\textit{Keywords---}} #1
}
\graphicspath{{}}
\DeclareMathAlphabet{\boldmathe}{T1}{cmr}{bx}{it}
\newcolumntype{L}[1]{>{\raggedright\arraybackslash}p{#1}}
\newcolumntype{C}[1]{>{\centering\arraybackslash}p{#1}}
\newcolumntype{R}[1]{>{\raggedleft\arraybackslash}p{#1}}
\newcommand{\mbf}[1]{\boldmathe{#1}}

\newcommand{\ft}[2]{{\textstyle\frac{#1}{#2}}}
\def\fdi{\slashed{\partial}}

\def\psib{\bar\psi}
\def\ha{\frac{1}{2}}
\def\id{\mathbbm{1}}
\def\D{\mathrm{d}}
\def\I{\mathrm{i}}
\def\tr{\mathrm{tr}}
\def\paslac{\partial^{\mathrm{slac}}}

\def\Nf{N_\mathrm{f}}
\def\Nr{\mathrm{N_\mathrm{r}}}
\def\Nfc{\mathrm{N_\mathrm{f}^\mathrm{crit}}}
\def\Nrc{\mathrm{N_\mathrm{r}^\mathrm{crit}}}

\def\vx{\mbf{x}}
\def\Z{\mathbbm Z}
\def\R{\mathbbm R}
\def\N{\mathbbm N}
\def\cC{{\mathcal C}}
\def\cD{{\mathcal D}}
\def\cF{{\mathcal F}}
\def\cL{{\mathcal L}}
\def\cO{{\mathcal O}}
\def\cP{{\mathcal P}}

\title{Symmetries of Thirring models on $3d$ 
lattices\footnote{Prepared for the 2021 Special Issue of the 
	journal “Symmetry” on “New Applications of Symmetry in Lattice Field Theory.”}
}
\author{Andreas W. Wipf \\ \texttt{wipf@tpi.uni-jena.de} \\[3mm]
	Julian J. Lenz\\ 
\texttt{julian.johannes.lenz@uni-jena.de}\\ [3mm]
Theoretisch-Physikalisches-Institut\\
	Friedrich-Schiller-Universit\"at, Max Wien Platz 1\\
	07743 Jena}
\date{\today}
\begin{document}
	\maketitle

\begin{abstract}
We review some recent developments
about strongly interacting relativistic Fermi theories 
in three spacetime dimensions. These models realize the
asymptotic safety scenario and are used to describe
the low-energy properties of Dirac materials in condensed 
matter physics. We begin with a general discussion
of the symmetries of multi-flavor Fermi systems
in arbitrary dimensions.
Then we review known results about the critical flavor 
number $N_\mathrm{crit}$ of Thirring 
models in three dimensions. Only models with flavor number below 
$N_\mathrm{crit}$ show a phase transition from a symmetry-broken
strong-coupling phase to a symmetric weak-coupling phase. 
Recent simulations with chiral fermions
show that $N_\mathrm{crit}$ is smaller than previously
extracted with various non-perturbative methods. Our simulations
with chiral SLAC fermions reveal that
for four-component flavors $N_\mathrm{crit}=0.80(4)$.
This means that all reducible Thirring models with $\Nr=1,2,3,\dots$
show no phase transition with order parameter.
Instead we discover footprints of phase transitions without order
parameter. These new transitions are probably smooth
and could be used to relate the lattice Thirring models to 
Thirring models in the continuum. For a single irreducible 
flavor, we provide previously unpublished values for the
critical couplings and critical exponents.
\end{abstract}

	\section{Introduction}
The Thirring model is a relativistic field theory for interacting 
fermions $\bar{\psi},\psi$. 
Its vector-vector interaction
\begin{equation}
\cL_\mathrm{int}=
\frac{g^2}{2}(\bar\psi\gamma^\mu\psi)^{2} = \frac{g^2}{2}J^{\mu}J_{\mu},
\label{intro1}
\end{equation}
establishes (after bosonization) a close relation to QED but it is also studied in various other contexts, e.g.\@ as a test bed for non-perturbative methods, as an example of asymptotic safety or as a toy model for chirally symmetric fermions.
The coupling constant $g^2$ has length-dimension $(d-2)$ and
is dimensionless in $2$ dimensions.

The $2$-dimensional massless model was introduced and
investigated by Walter 
Thirring in 1958 \cite{thirring} and represents an exactly 
solvable conformal field theory with analytically
known $n$-point correlation functions \cite{chh4tk68,thirring_sachs}.
The massive model can be solved with the Bethe ansatz
which yields the mass spectrum and scattering matrix elements.
In higher dimensions the model in not soluble and not
renormalizable in perturbation theory.  But it is renormalizable beyond
perturbation theory -- above $2$ and below $4$ dimensions
it is the prototype of an asymptotically safe theory
\cite{gies-janssen}.

We begin with discussing the symmetries of
Thirring models with Lagrangian
 \begin{equation}
 \cL=\cL_m+\frac{g^2}{2\Nf} J^\mu J_\mu,\quad
 \cL_m=\sum_a \psib_a(\fdi+m)\psi_a \,.
 \label{intro3}
 \end{equation}
for $\Nf$ flavors of fermions $\psi_1,\dots,\psi_{\Nf}$ 
in arbitrary dimensions. 
Then we shall discretize
the models on a (euclidean) spacetime lattice 
with chiral fermions keeping all continuum symmetries
besides Poincare invariance. Finally we shall focus
on the lattice models in $3$ dimensions
and discuss the possible breaking of symmetries,
depending on the number of flavors and the
interaction strength. Finally we summarize the present
status concerning the critical flavor number $\Nfc$
which separates the systems with spontaneous 
parity breaking from those without symmetry breaking.

An irreducible spinor in $d$-dimensions has $d_s=2^{\lfloor d/2\rfloor}$
components, where $\lfloor a\rfloor$  is the largest integer less or equal
to $a$. In euclidean space we can and will always 
choose hermitian $\gamma^\mu$ matrices.
In even dimensions there exists one irreducible representation
of the Clifford algebra, whereas in odd dimensions
there are two. The hermitean matrix
\begin{equation}
\gamma_*=-\I^{\lfloor d/2\rfloor}\gamma^0\cdots \gamma^{d-1},
\quad\gamma_*^2=\id\label{intro5}
\end{equation}
generalizes $\gamma_5$ to arbitrary dimensions.
In even dimensions it anticommutes with the $\gamma^\mu$ and
in odd dimensions it commutes with the $\gamma^\mu$ and is $\id$ in
one irreducible representation and $-\id$ in the other irreducible
representation. The bilinear fields
\begin{equation}
S=\sum_{a=1}^{\Nf}\bar{\psi}_a\psi_a,\quad
P=\I \sum_{a=1}^{\Nf}\bar{\psi}_a\gamma_*\psi_a,\quad 
J^\mu=\sum_{a=1}^{\Nf}\psib_a\gamma^\mu\psi_a
\label{intro7}
\end{equation}
are of particular interest here:
the current density $J^\mu$ enters the
Lagrangian of the Thirring models  and $S,P$
may or may not condense in an equilibrium
state. One should note that $S$ and $P$ are
not independent in odd dimensions because
$\gamma_*\propto\id$ is trivial.

\section{Symmetries of Fermi systems}
\label{sec:symmetries}
The symmetries of Thirring models with Lagrangians (\ref{intro3})
are the usual (Euclidean) spacetime 
symmetries (including parity), chiral rotations and charge conjugation.
\paragraph{Charge conjugation:}
The transformation of spinor fields and $\gamma^\mu$-matrices under charge conjugation
\begin{equation}
\psi_c=\cC\psi^*,\quad
\gamma_\mu^T=\eta_c\, \cC^{-1}\gamma_\mu\cC,\quad\eta_c\in\{-1,1\}\,,
\label{symm7}
\end{equation}
are used to investigate the sign problem in four-Fermi theories.
Actually there are two matrices $\cC$ with $\eta_c=\pm 1$ in even dimensions, 
one $\cC$ with $\eta_c=-1$ in $3+4n$ dimensions and one $\cC$
with $\eta_c=1$ in $1+4n$ dimensions for $n\in\N_{0}$ \cite{wipf_book}.
\paragraph{Parity:} A parity transformation flips the sign of a spatial coordinate,
e.g.
\begin{equation}
x\mapsto x'=Px,\quad P=(P^\mu_{\;\,\nu})
=\hbox{diag}(1,1,\dots,1,-1)\,.\label{symm9}
\end{equation}
Scalar and pseudo-scalar fields have even and odd parity,
respectively.
A spinor and its conjugate transform according to
\begin{equation} 
\psi(x)\mapsto\psi_P(x')=\cP\psi(x),\quad 
\psib(x)\mapsto\psib_P(x')=\alpha \psib(x)\cP^{-1}\,,
\label{symm11}
\end{equation}
where the parity matrix $\cP$ satisfies
\begin{equation}
\cP^{-1}\gamma^\mu \cP=\alpha P^\mu_{\;\,\nu}\gamma_\nu
\Longrightarrow \cP^{-1}\gamma_*\cP=-\alpha^d\gamma_*,
\quad \alpha^2=1\,.
\label{symm13}
\end{equation}
The density $\cL_{m=0}$ is parity invariant,
$J^\mu$ is a vector and $P$ is a pseudoscalar,
\begin{equation}
	P_P(x')=-P(x)\,.\label{symm15}
\end{equation}
The transformation properties of $S$ depend on the
dimension, i.e.\@
\begin{equation}
	S_P(x')=\alpha S(x),
\end{equation}
such that $S$ is a scalar in even dimensions while
$S\propto P$ is pseudoscalar in odd dimensions as detailed below.

\subsection{Even dimension}
In even dimensions there is only one irreducible representation of
the Clifford algebra and there exists a canonical choice
for the free Lagrangian $\cL_m$ with the same $\fdi$ acting on
all flavors.
For a vanishing mass the Dirac operator $\fdi$ 
anti-commutes with $\gamma_*$ and we
may rotate the chiral (left- and right-handed) fermions 
\begin{equation}
\psi_{\pm}=P_\pm \psi=\ha(\id\pm\gamma_*)\,\psi\,,\label{symm19}
\end{equation}
independently among each other,
\begin{equation}
\psi_+\mapsto U_+\psi_+,\quad \psi_-\mapsto U_-\psi_-\quad
U_+,U_-\in \hbox{U}(\Nf)\,.\label{symm21}
\end{equation}
These chiral rotations leave the Lagrangian of
massless fermions $\cL_{m=0}$ invariant. The bilinear
$\propto\psib\psi$ is only invariant under the
diagonal subgroup with $U_+=U_-$. Thus, a mass term or 
a condensate $\langle\psib\psi\rangle$ break
the chiral symmetry explicitly respectively spontaneously 
to the vector flavor symmetry $\text{U}_V(\Nf)$.

In even dimensions there is a parity matrix $\cP$ with
$\alpha=1$ such that the bilinear $S$ is a scalar and $P$ 
a pseudoscalar, see (\ref{symm15}).
In addition, there exist two matrices
$\cC$ which obey (\ref{symm7}), one for each sign of $\eta_c$.

\subsection{Odd dimensions}
In odd dimensions there is no notion of chirality in an irreducible
representation and there are only the vector flavor rotations
\begin{equation}
\psi\mapsto U\psi,\quad \psib\to\psib U^\dagger,\quad U\in \text{U}(\Nf)\,,\label{symm25}
\end{equation}
which leave $\cL_m$ in (\ref{intro3}) invariant.
The bilinears (\ref{intro7}) are singlets under these rotations.
In odd dimensions there exists a parity matrix $\cP$ which fulfills (\ref{symm13}) with $\alpha=-1$ such that  the bilinear $S$ 
is parity-odd \cite{wipf_book},
\begin{equation}
S_P(x')=-S(x)\,.
\label{symm31}
\end{equation}
Actually, the bilinears  $S$ and $P$ are not independent, $S=\pm \I P$, since  $\gamma_*=\pm\id$ in the two irreducible representations of 
the Clifford algebra.
We see that a mass term or bilinear condensate
break the $\Z_2$-parity symmetry explicitly or spontaneously.

The last statement applies to systems with odd $\Nf$ only,
as for even $\Nf$ we can build a parity invariant massive Lagrangian. 
For example, for $\Nf=2$ one combines the two irreducible flavors
to one reducible flavor and acts with the inequivalent irreducible representations $\gamma^\mu$ 
and $-\gamma^\mu$  on the upper and lower components,
\begin{equation}
\cL_m=\bar\Psi(\Gamma^\mu\partial_\mu+m)\Psi,\quad 
\Psi=\begin{pmatrix}\psi_1\\ \psi_2\end{pmatrix},\quad
\Gamma^\mu=\sigma_3\otimes\gamma^\mu
\,.\label{symm33}
\end{equation} 
For the reducible system $\tilde \cP=\I\sigma_2\otimes\cP$
is a parity matrix satisfying the defining relation 
(\ref{symm13}) with $\Gamma^\mu$ 
and $\alpha=1$ such that $\bar\Psi\Psi$ and hence $\cL_m$ in (\ref{symm33}) are parity even. This construction straightforwardly 
generalizes to an even flavor number $\Nf=2\Nr$. One just groups the 
$2\Nr$ irreducible 
flavors into $\Nr$ reducible flavors  $\Psi_1,\dots,\Psi_{\Nr}$.
The Lagrangian for the latter reads\footnote{we also rescale
the coupling such that $g^2/2\Nf\to g^2/2\Nr$.}
\begin{equation}
\cL_m
=\sum_{a=1}^{\Nr}\bar\Psi_a(\Gamma^\mu\partial_\mu+m)\Psi_a
+\frac{g^2}{2\Nr}J^\mu J_\mu,\quad 
J^\mu=\sum_a\bar\Psi_a\Gamma^\mu\Psi_a\,.
\label{symm35}
\end{equation}
By construction this parity invariant reducible model 
is invariant  under $\text{U}(\Nf)=\text{U}(2\Nr)$ rotations. 
$\cL_m$ can be obtained by a dimensional 
reduction of the 
Thirring model in one dimension higher.
The various symmetries of reducible four-Fermi systems
are well explained in \cite{gies-janssen,gehring-gies}.

\subsection{Fierz-identities}
It may happen that two seemingly different looking 
four-Fermi theories are equivalent on account of 
Fierz-identities. In $2$ and $3$ dimensions irreducible 
spinors have $2$ components only and there is a direct way
to relate different one-flavor models.
Indeed, for a $2$-component anti-commuting $\psi$ 
an arbitrary four-Fermi term $(\psib A\psi)^2$ 
at a fixed spacetime point must
be proportional to $\psib_1\psi_1\psib_2\psi_2$ and
\begin{equation}
(\psib A\psi)^2=
\det A\,(\psib\psi)^2,\quad \Nf=1,\;\; d =2,3\,.\label{symm39}
\end{equation}
\begin{enumerate}
\item \emph{One-flavor models in two and three dimensions:}
The hermitian $2\times 2$ matrices $\gamma^\mu$ 
have determinant $-1$ and we conclude
\begin{equation}
\frac{g^2}{2}(\psib\gamma^\mu\psi)^2=
-\frac{d\,g^2}{2}(\psib\psi)^2,\quad \Nf=1,\;\; d=2,3\,.\label{symm41}
\end{equation}
It follows that the one-flavor Thirring model is equivalent to the 
one-flavor Gross-Neveu (GN) model. Thus the latter is not only invariant under
$\text{U}(1)\times \Z_2$ but also under the larger symmetry group
$\hbox{U}_V(1)\times \hbox{U}_A(1)$.
\item A comparable simple relation does not exist for $\Nf>1$
or $d>3$. For example, the general Fierz identity in $3$ dimensions
implies that for $\Nf$ irreducible flavors
\begin{equation}
\frac{g^2}{2}(\psib\gamma^\mu\psi)^2=-\frac{g^2}{2\Nf}
(\psib\psi)^2-\frac{g^2}{\Nf}
\sum_{a,b}\,(\psib^a\psi^b)(\psib^b\psi^a),\quad
d=3\,.\label{symm43}
\end{equation}
This means that the Thirring interaction is converted into a GN
interaction plus a tensor-tensor coupling.
\end{enumerate}
\subsection{Hubbard-Stratonovich transformation}\label{sec:hubbard}
It is possible to eliminate the four-Fermi terms in the
Lagrangian by a Hubbard-Stratonovich transformation with the help
of an auxiliary vector field $v_\mu$,
\begin{equation}
\cL=\bar\psi\cD_m\psi
+\frac{\Nf}{2g^2} v_\mu v^\mu\,.\label{symm51}
\end{equation}
The Dirac operator contains the auxiliary  field,
\begin{equation}
\cD_m=\gamma^\mu(\partial_\mu+\I v_\mu)+m\,.
\label{symm53}
\end{equation}
The classical systems with Lagrangians (\ref{symm51}) 
and (\ref{intro3}) are equivalent as follows from
the field equation for the auxiliary 
vector field. The equivalence also holds for the 
quantized system, since $v_\mu$ is non-dynamical,
enters the Lagrangian at most quadratically and thus
can be integrated over explicitly in the path integral. 

In passing we note that the equivalent one-flavor GN-model with
interaction term (\ref{symm41}) in $d=2,3$ can be bosonized with
a scalar field $\sigma$ as
\begin{equation}
\cL=\bar\psi(\fdi+m+\sigma)\psi+\frac{1}{2dg^2}\sigma^2
\,.\label{symm55}
\end{equation}
Interestingly, the fermion determinant of $\fdi+\sigma$ 
is (generically) complex, whereas that of $\cD_0$ 
is real. This means
that by a Fierz-reshuffling from the scalar into the vector
channel one can soften or even solve the ubiquitous sign
problem.

\section{Critical Flavor Number of 3d Thirring models}
In the following we focus on the
$3$-dimensional multi-flavor Thirring model. The irreducible
systems have Lagrangian (\ref{intro3}) for a multiplet
$\psi$ of $\Nf$ two-component spinor fields
and the reducible ones have Lagrangian (\ref{symm35})
for a multiplet of $\Nr$ four-component 
reducible fields.
There are several reasons for considering the 
subclass of reducible Thirring models. 
First and most important, most applications in condensed
matter physics deal with Dirac-type materials where naturally
one is lead to the parity-invariant reducible models. 
The most prominent example is of course graphene, where low-energy
electronic excitations exhibit a linear dispersion around
two Dirac points in the first Brillouin zone.
Second, most results in the literature are obtained for 
the reducible systems, and for an ease of comparison
we consider such systems as well. The reducible systems
are obtained by a torus-reduction of four-dimensional
interacting Fermi systems.

The reducible models have no sign problem. For example,
for one reducible flavor $\Psi$ the fermion determinant of $\cD_m^\mathrm{red}$ in
\begin{equation}
\cL_\mathrm{red}=\bar\Psi \cD^\mathrm{red}_m\Psi
+\frac{1}{2g^2} v_\mu v^\mu,\quad 
\cD^\mathrm{red}_m=\begin{pmatrix}\cD_m&0\\ 0 &-\cD_{-m}
\end{pmatrix}
\label{cfn1}
\end{equation}
is $\det(m^2-\cD_0^2)>0$, since
$\cD_0=\cD_{m=0}$ defined in (\ref{symm53}) is
anti-hermitean.
Note that the massive reducible model with $\Nr$ flavors is 
not equivalent to the irreducible model with $2\Nr$ flavors.
The two irreducible flavors have opposite mass and this
explains why there is no breaking of parity in the reducible model.
Only in the limit $m\to 0$ are the two models
equivalent. The passage from one to the other involves a relative
rotations of the two irreducible flavors which combine
to a reducible flavor such that
\begin{equation}
\cD_0^\mathrm{red}
\mapsto\begin{pmatrix}\cD_{0}&0\\ 0 &\cD_{0}
\end{pmatrix}\,.
\label{pas1}
\end{equation}
At the same time the chiral condensate of one formulation
transforms into the staggered condensate in the other
\begin{equation}
\langle \bar\Psi\Psi\rangle=\langle\bar\psi_1\psi_1\rangle-
\langle\bar\psi_2\psi_2\rangle\,,\label{pas3}
\end{equation}
and vice versa.

\subsection{Small and large-\texorpdfstring{$\Nf$}{} limit}
Integrating over the fermion fields in the (euclidean)
functional integral with Lagrangian (\ref{symm51}) yields
\begin{equation}
Z=\int\cD v_\mu\,
e^{-\Nf S_\mathrm{eff}[v_\mu]},\quad
S_\mathrm{eff}=\frac{1}{2g^2}\int \D^3x\,v_\mu^2-\ha\log\det(-\cD_m^2)\,.
\label{cfn3}
\end{equation}
In the large-$\Nf$ limit the absolute minimum of 
$S_\mathrm{eff}$ dominates the path 
integral such that the free energy per flavor
simplifies considerably,
\begin{equation}
F=-\frac{1}{\beta\Nf}\log Z
\stackrel{\Nf\to\infty}{\longrightarrow}
\frac{1}{\beta}\min_{v_\mu} S_\mathrm{eff}[v_\mu]\,.\label{cfn5}
\end{equation}
The Euler-Lagrange equation for $S_\mathrm{eff}$ is just
the gap equation which determines the minimizing field $v_\mu$. 
For a translation-invariant equilibrium
state the minimizing field is homogeneous and for a constant $v^\mu$
the eigenvalues of $\cD_0$ come in pairs $\pm\lambda$ which implies
\begin{equation}
\det(-\cD_m^2)=\det(\cD^\mathrm{red}_m)\,.\label{cfn6}
\end{equation}
This means that in the large-$\Nf$ limit the irreducible
and the parity invariant reducible systems 
are identical, or that the irreducible models do not break 
parity. This observation is
supported by an explicit calculation
of the free energy density (effective potential)
\begin{equation}
	U_\mathrm{eff}=\frac{F}{V}=
	\frac{1}{2g^2_\mathrm{ren}}v_\mu v^\mu
	+U_\mathrm{\,free}(T,m^2),
	\quad g^2=\frac{4\pi g_\mathrm{ren}^2}{4\pi+\Lambda g_\mathrm{ren}^2}\,,
	\label{cfn7}
\end{equation}
where $\Lambda$ is the momentum cutoff.
Besides the free energy density of the free Fermi gas one
only gets a renormalization of the Thirring coupling.
The parity condensate is obtained by differentiating $U_\mathrm{eff}$
with respect to the trigger mass $m$.
Since the derivative does not depend on the auxiliary field we
conclude
\begin{equation}
\langle\psib \psi\rangle
=\frac{\partial}{\partial m}\,U_\mathrm{free}
\stackrel{m\to 0}{\longrightarrow}0,\quad 
\Nf\to\infty\,.\label{cfn9}
\end{equation} 
Thus there is no parity condensate for a large number of
flavors. 

In the other limit $\Nf=1$ the Thirring model
is equivalent to the $3$-dimensional Gross-Neveu model, 
see section \ref{sec:hubbard}, and the latter
shows spontaneous breaking of parity \cite{hoefling}.
Thus we have
\begin{equation}
\langle\bar\psi\psi\rangle\neq 0,\quad \Nf=1\,.\label{cfn11}
\end{equation}
Since parity is broken for $\Nf=1$
and unbroken for $\Nf\to\infty$ we must conclude that 
there exists a critical flavor number $\Nfc$ separating
the systems with symmetry breaking from those without 
symmetry breaking.

For reasons explained above the parity invariant
reducible models are of particular interest. 
In the limit $\Nf\to\infty$
they are identical to the irreducible Thirring models.
Since the parity breaking model with one irreducible
flavor is not in the class 
of reducible models - formally it has $0.5$ reducible flavors --
there is no compelling argument that there
must exist a critical flavor number $\Nrc$ within the class of 
reducible models.

First investigations of Thirring models 
with Schwinger-Dyson equations, partly in
combination with a large-$\Nf$ expansion, date back to the 
nineties of the last century.
In Table~\ref{tab:ncrit_past} we collected
values for the critical flavor numbers $\Nfc$ and $\Nrc$
obtained with Schwinger-Dyson (SD) equations or expansions
in $1/\Nf$, a Gaussian ansatz for the
state of interest in the Schr\"odinger picture,
the functional renormalization group (FRG) and dedicated 
lattice simulations.

\begin{table}[h] \centering
	\centering
	\caption{Critical flavor numbers $\Nfc$ and $\Nrc$.
	SD means Schwinger-Dyson and FRG functional
	renormalization group. For example, SD-equations predicted
	that for $\Nf=1,2,3$ there is a parity condensate $\langle\psib\psi\rangle$ and
	simulations with staggered fermions that for $\Nr=1,2,\dots,6\,$ there 
	is a condensate $\langle\bar\Psi\Psi\rangle$.\label{tab:ncrit_past}}	
	\begin{tabular}{m{5.5cm} m{2cm} m{2.2cm} r r }\toprule
			method&$\Nfc$&$\Nrc$&references&years\\ 
		\midrule
			SD equations&$6.48$&$3.24$&\cite{Gom91}&1991\\ 
		&$\infty$  & &\cite{HP94}&1994\\
		&&$4.32$&\cite{Ito95,Sug97}&1995, 1997\\ \midrule
		$1/\Nf$-expansion&&$2.00$ &\cite{Kon95}&1995\\
		&$<3$&&\cite{AP98}&1998\\ \hline
		Gaussian approximation&$\infty$&$\infty$&\cite{HLY94}&1994\\ \midrule
		FRG&&$5.1$&\cite{JG12}&2012\\ 
		&&$\lessapprox2$&\cite{DGK19}&2019\\ \midrule
		lattice (staggered)&&$(4\dots 6)$&\cite{DHM97,DH99}&1997, 1999\\  
		&&$6.61$&\cite{HL99,CHS07}&1999, 2007\\  \midrule
		lattice (slac)&$\lessapprox 9$ odd&$0.80$&\cite{WSW17,LWW19} &2017, 2019\\ \midrule 
		lattice (domain wall)&&$(1\dots 2)$ &\cite{SH18,HMW20} &2018, 2020 \\ 
		\bottomrule
	\end{tabular}
\end{table}
Early lattice studies were performed with 
light staggered fermions to
recover the chiral symmetry in the continuum limit.
With the help of an HMC algorithm, 
simulations with an even $\Nr$ and subsequently
with non-integer $\Nr$ have been presented
in \cite{DH99,HL99}.
In a subsequent lattice study \cite{CHS07} with a similar 
setup, the authors concluded that the critical 
flavor number is $\Nrc=6.6(1)$.

More recent analytic studies as well as simulations 
with massless SLAC fermions yield different results -- they 
favor smaller values of $\Nrc$. Lattice models with
these chiral fermions have the same internal symmetries as the
continuum models. It was
demonstrated, that the U($2\Nr$) symmetry of the \emph{reducible
model} is never broken 
for any integer number of $4$-component flavors \cite{WSW17}.
In a subsequent publication the critical flavor
number $\Nrc= 0.80(4)$ has been calculated \cite{LWW19}. 
\emph{Irreducible Thirring models} with an odd number of flavors 
behave differently from the reducible models.
The former show a parity broken phase for $\Nf\lessapprox 9$.

Independent simulations with $4$-component domain-wall 
fermions (DWF),
in which one adds an extra dimension to the Dirac operator,
pointed to a critical flavor number $\Nrc$ 
below $2$ \cite{SH18} . In a follow-up publication with DWF
it was demonstrated, that the model with $\Nr=1$
shows a phase transition with order parameter \cite{HMW20},
implying that $1< \Nrc<2$.
The discrepancy of the results obtained
with SLAC and DW fermions may be due to uncertainties
in the extrapolation to an infinite domain-wall separation.
In \cite{HMW20} it was speculated that the two lattice
approaches describe different continuum theories, and that the bulk
DWF formulation more closely conforms to 
a picture of the strong dynamics in which the auxiliary 
vector field resembles a gauge field.

\section{Lattice simulations with chiral SLAC derivative}
In our simulations we use the chiral and non-local
SLAC-derivative on a hypercubic lattice. 
The SLAC fermions have been used with great
success to
\begin{enumerate}
\item calculate the critical coupling of $\phi^4_2$-theory to high 
precision \cite{WW14},
\item obtain an accurate value for the 
step scaling function in the two-dimen\-sional 
nonlinear O(3) model \cite{FKWW12},
\item 
calculate Ward-identities, the ground state structure, 
low-lying masses and
the breaking or restoration of supersymmetry in
low-dimensional supersymmetric 
Wess-Zumino models \cite{BKUW08,KBUWW08,WW14},
\item find accurate values for the
critical exponents of GN model \cite{dissschmidt,LL19},
\item discover inhomogeneous structures 
in the multi-flavor $\Z_2$ and U$(1)$ Gross-Neveu models \cite{LPWWW20,LMW21}.
\end{enumerate}
Lattice models with SLAC fermions have 
various advantages. In the present context
the most relevant one is that they inherit all global
inner symmetries and discrete space-times symmetries
of the continuum models. For example, the reducible 
lattice Thirring model in three dimensions 
is invariant under U$(2\Nr)$ chiral transformations and $\Z_2$-parity.
The lattice derivative $\paslac_\mu$ is anti-hermitean
such that $\I\fdi$ is hermitean. This property
is used to prove that certain fermion operators have no
sign problem. Furthermore, the auxiliary field $v_\mu$ 
in $\cD_m$ is a non-compact site variable and not a 
link variable as in some other lattice formulations.
A further advance is that the Dirac operator has no 
doublers, and no rooting is necessary to describe
systems with a small flavor number.
Last but not least SLAC lattice fermions are cheap compared 
to local chiral fermions. 

It is well-known that the non-local SLAC derivative
leads to problems in lattice gauge theories \cite{KS79}.
Indeed, when the ordinary lattice derivative $\partial$ 
is replaced by a covariant derivative $D$, then one can 
perform a local gauge transformation which does not 
change the action but sends the canonical momentum $p=\I \partial$ 
to the edge of the Briolloin zone, where $p$ jumps. 
In case a complete gauge fixing is available (because
of the Gribov-problem such a fixing will not be continuous) the
problem with the discontinuous dispersion relation
may be overcome. This has been demonstrated in lower-dimensional 
supersymmetric gauge models \cite{Hanada07}.

\subsection[SLAC derivative]{SLAC derivative\footnote{In the
published version (Symmetry \textbf{14} (2022) 333)
the SLAC derivative for anti-periodic BC is incorret.}}
To find the SLAC derivative on a
finite lattice one first Fourier-transforms a wave function,
multiplies the transformed wave function with the
momentum and transforms back to coordinate space.
In the spatial directions we impose \emph{periodic boundary
conditions} for which the momenta $p_\ell$ are from $2\pi\Z/N_s$. 
In the time-direction we impose \emph{anti-periodic
boundary conditions} for which the momenta are from
$2\pi (\Z+\ha)/N_t$.
We choose the $N$ momenta $p_\ell$ symmetric to the origin,
such that the
edge of the Brillouin zone has maximal distance from
the pole of the propagator. As a result we need an odd number
of momenta and hence an odd number $N_s$
of lattice points in the spatial directions, and
an even number $N_t$ of lattice points in the time-direction,
\begin{align}
p&\in \{p_\ell\vert \ell=1,\dots,N\},\quad \;
p_\ell=\ft{2\pi}{N}\big(\ell-\ft{N+1}{2}\big)\,,
\\
x&\in\{x_k\vert k=1,\dots,N\},\quad 
x_k=\ft{2\pi}{N}\big(k-\ft{N+1}{2}\big)
\,,\label{slac1}
\end{align}
with $N=N_s$ odd in the space-directions and $N=N_t$ even
in the time-direction.
The anti-symmetric and real lattice derivative in a given
direction takes the form
\begin{equation}
\langle x\vert\partial^\text{slac}\vert x'\rangle=
\partial^\text{slac}_{x-x'} 
	= \begin{cases}
		0 &\hskip3.3mm x=x'\,,\\
		\frac{\pi}{N_s}\frac{(-)^{k-k'}}
		{\sin(x-x')/2} &\hskip3.3mm x\neq x'\,,
	\end{cases}\label{slac3}
\end{equation}
where $x=x_k$ and $x'=x_{k'}$ are sites on the lattice defined
in (\ref{slac1}).

For finite-temperature systems in $2$ spatial dimensions
the SLAC derivatives $\paslac_\mu$ take the form
\begin{equation} 
\paslac_{0,\xi}=\partial^\text{slac}_{\xi_0}\delta_{\xi_1}\delta_{\xi_2}
,\quad
\paslac_{1,\xi}=\delta_{\xi_0}\partial^\text{slac}_{\xi_1}\delta_{\xi_2}
,\quad
\paslac_{2,\xi}=\delta_{\xi_0}\delta_{\xi_1}\partial^\text{slac}_{\xi_2}
\,,\label{slac9}
\end{equation}
where we abbreviated $\delta_{\xi_1}\equiv \delta_{0\xi_1}$.

Below we shall present the results of simulations at low temperature
on a $N\times (N-1)^2$ lattices with $N=8,12,16,24$. In the
simulations we used pseudo-fermions and the rational HMC-algorithm 
with operator
\begin{equation}
\Big(\det (\cD_0^\dagger\cD_0)^{\Nf/2N_\mathrm{PF}}\Big)^{N_\mathrm{PF}},\quad
N_\mathrm{PF}\approx 2 \Nf\,.\label{slac11}
\end{equation}
The inverse of the shifted operator enters the rational 
approximation based on a multi-mass conjugate
gradient (CG) solver.  During the CG iterations 
the derivative is applied many times to a pseudo-fermion field.
Thereby we make use of a special property of the SLAC derivative:
It is diagonal in momentum space, such that
\begin{equation}
(\cD_m\psi)(x)=\cF^{-1}\Big[\I\slashed{p}\cF[\psi](p)\Big](x)
+\big(\I\gamma^\mu v_\mu(x)+m\big)\psi(x)\,,
\end{equation}
where $\cF$ denotes the Fourier transformation.
Instead of using a three-dimensional (parallelized) Fourier 
transformation, we apply one-dimensional Fourier transformations
that are computed in parallel.

To estimate fermion propagators we used $N_\mathrm{est}\approx 
200\Nf$ stochastic estimators. For most measurements we generated
approximately $5000$ configurations. We estimated 
the finite size corrections and checked that they are
under control.

For the models with $\Nr\in [0.5,1.1]$ we used the
parity-even extensions to non-integer $\Nr$. More
precisely, after integrating out the fermions we
arrive at the effective action (\ref{cfn3}) in which
$\Nr$ is only a prefactor. We then continue $\N_r\in\N$ 
to real values. This
formal procedure is similar to other studies where
$\Nr$ only appears as a parameter but we should
stress that for $\Nr\notin\N$ such a definition most likely does not
describe a local quantum field theory in the continuum.
In particular, reducible models with half-integer $\Nr$ 
obtained via this procedure need not be equivalent to
irreducible models with integer $\Nf=2\Nr$.

\section{Dual formulation and effective potentials}
Similarly as in the fermion bag algorithm of Chandrasekharan \cite{Chandrasekharan10},
one can integrate out the interaction part of the partition sum 
to obtain a formulation in terms of new spin like
variables $k_{x\,i}^{ab} \in \{0,1\}$, where
$i$ relates to the spinor degree of freedom and $a,b$ to the
flavor degree of freedom \cite{dissschmidt,lenz18}. For non-local SLAC-fermions the dual formulation is a bit more involved
than for ultra-local staggered fermions. Actually, 
it is advantageous to
use the Fierz-identity (\ref{symm43}) and bosonize the
resulting four-Fermi theory. For $m=0$ this yields the tensor-scalar formulation
\begin{equation}
\cL=\bar\Psi (\I\fdi+\I T+\I\phi)\Psi+
\frac{\Nf}{4g^2}\,\tr\, T^2+\frac{\Nf}{2g^2}\phi^2,\quad T^\dagger=T,\;\phi\in\R\,,\label{tm1}
\end{equation}
which is equivalent to the vector formulation of the Thirring model.
Differentiating the Boltzmann factor with respect to 
the components of the tensor field $T_{ab}$ yields the Dyson-Schwinger equations
\begin{equation} 
\langle T_{ab}\rangle\propto \langle \psib_a\psi_b\rangle\,.\label{tm3}
\end{equation}
Since $T$ transforms under chiral rotations as $T\to U T U^\dagger$, 
the expectation value $\langle T_{ab}\rangle$ serves as
an order parameter for chiral symmetry. 
In the scalar-tensor formulation we can probe for 
condensates in all channels represented by $T_{ab}$. 
For example a parity breaking condensate, a
chirality breaking condensate 
or a non-time-reversal invariant Haldane-term \cite{FH88}. 
Unfortunately this formulation has a severe sign problem 
and cannot be used directly in simulations. But exploiting
its dual formulation we are able to express various
quantities of interest, for example coefficients in the
expansion of the effective potential, to expectation values
in the sign-problem free vector formulation.

\subsection{Effective potential versus condensates}\label{subsec1}
The expectation value $\langle T_{ab}\rangle$ can be diagonalized
by a chiral rotation and it suffices
to calculate the effective potential (denoted by $V$ to distinguish
it from the effective potential in the vector formulation) on diagonal
order parameters $\langle T\rangle=t_i H^i$, where 
$\{H^i\}$ forms a basis of the space of diagonal hermitean matrices,
\begin{equation} 	
V(\langle T_{ab}\rangle)=
V(t)=-\frac{1}{V}\log \sum \limits_{n=0}^{2 \Nf} \sum \limits_{i=1}^{\Nf} a_{n,i} \left(t_i\right)^n\,.\label{tm5}
\end{equation}
With the help of the dual formulation one proves that the coefficients 
$a_{n,i}$ are given by expectation values of powers of fermion bilinears
and these expectation values can be calculated in the vector formulation.

\begin{figure}[h!]
	\includegraphics[width=\linewidth]{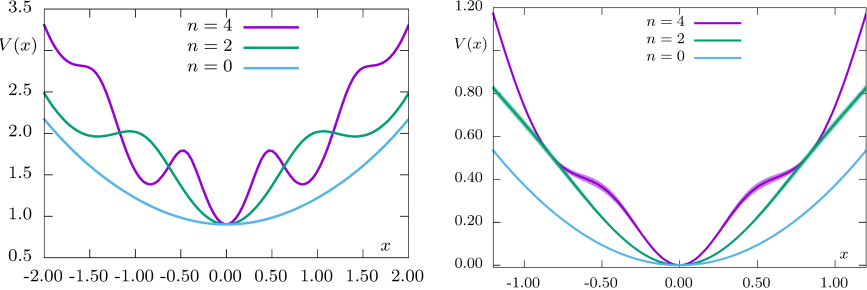}
	\caption{Effective potential for the three channels of the
		Thirring model with $\Nr=2$ or equivalently $\Nf=4$ flavors. Left panel:
		strong coupling limit. Right panel: simulation 
		on a lattice with $N=16$ and inverse coupling $\lambda=0.118$.
	Figures taken from \cite{WSW17}	\label{fig:effpotN4}.}
\end{figure}

The minima of $V$ are attained for fields
\begin{equation}
\langle T\rangle_\text{min}
=\frac{2x}{\Nf}\,\mathrm{diag}\big( 
\underbrace{1,\dots,1}_{n_+},\underbrace{-1,\dots,-1}_{n_-}\big)
,\quad n_+\geq n_-\,,\label{tm11}
\end{equation}
where permutations of the diagonal elements 
lead to equivalent minima.
The physically distinct ones are characterized by the amplitude 
$x$ and by
\begin{equation}
n=n_+-n_-=
\begin{cases} 0,2,4, \dots,\Nf, & \text{for }\Nf \text{ even\,,} \\ 
1,3,5, \dots,\Nf & \text{for }\Nf, \text{ odd\,.} \end{cases}\label{tm13}
\end{equation}
An order parameter (\ref{tm11}) with $x>0$ gives rise to
the symmetry breaking pattern
\begin{equation}
U(\Nf)\otimes \Z_2\to U(n_+)\otimes U(n_-)\,,\label{tm14}
\end{equation}
which is different for an odd and an even number
of irreducible flavors. For example, for three and four
flavors we find
\begin{align}			
		\Nf&=4:\quad (n,n_+,n_-)\in\{(0,2,2),\;(2,3,1),\;(4,4,0)\}\,,\label{tm15a}\\
		\Nf&=5:\quad (n,n_+,n_-)\in\{(1,3,2),\;(3,4,1),\;(5,5,0)\}\,.
		\label{tm15b}
\end{align}
Only reducible models with even $\Nf=2\Nr$
permit a symmetry breaking with $n=0$, in which
case $\text{U}(2\Nr)$ breaks to $\text{U}(\Nr)\times \text{U}(\Nr)$.
This breaking is induced by a staggered condensate $\sum(-1)^a\bar{\psi}_a\psi_a$, which
corresponds to a chiral condensate $\bar{\Psi}\Psi$ (and not the Haldane
mass) in the reducible formulation, see eq. (\ref{pas3}).
In this channel the broken system is parity invariant.
In Figure \ref{fig:effpotN4} the potentials for
$\Nf=4$ and the three feasible breaking patterns in (\ref{tm15a})
are depicted.
The left panel shows the analytic results in the strong coupling 
limit whereas the right panel shows the results of simulations on a $16\times 15\times 15$-lattice with inverse coupling $\lambda=\Nf/2g^2=0.118$.

We see that in all three channels the minimum of the
effective potential is at $\langle T\rangle=0$.
Thus we do not observe spontaneous symmetry breaking (SSB)
for the given $\lambda$ and on the chosen lattice.
With increasing inverse coupling $\lambda$ the minima at the origin
get more pronounced and the only way to see SSB
is to decrease $\lambda$. We shall
see below, that for all admitted values of $\lambda$ we see
no SSB in the reducible models.

\begin{figure}[h]
	\includegraphics[width=\linewidth]{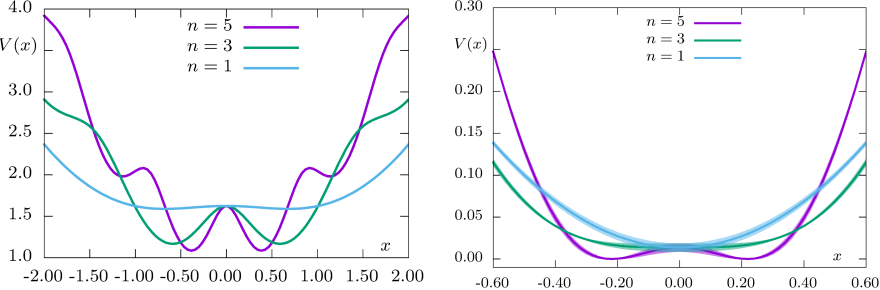}
	\caption{Effective potential for the three channels of the
Thirring model with $\Nf=5$ irreducible flavors. Left panel:
		strong coupling limit. Right panel: simulation 
		on a lattice with $N=16$ and inverse coupling $\lambda=0.102$.
		Figures taken from \cite{WSW17}.\label{fig:effpotN5}}
\end{figure}

For an odd number of irreducible flavors the situation
looks differently. Figure \ref{fig:effpotN5} shows the effective
potentials in the three channels (\ref{tm15b})
available for $\Nf=5$.
Again the result in the strong coupling limit is shown
in the left panel and the simulation results for $\lambda=0.102$ in
the right panel. We observe that a condensate $\langle T\rangle$ with
$n=5$ forms. This condensate does not break the chiral symmetry U$(5)$ 
but it breaks parity. If one decreases the coupling $g^2$ then one
observes a transition into the symmetric phase.
 
\begin{figure}[h]
 	\includegraphics[width=\linewidth]{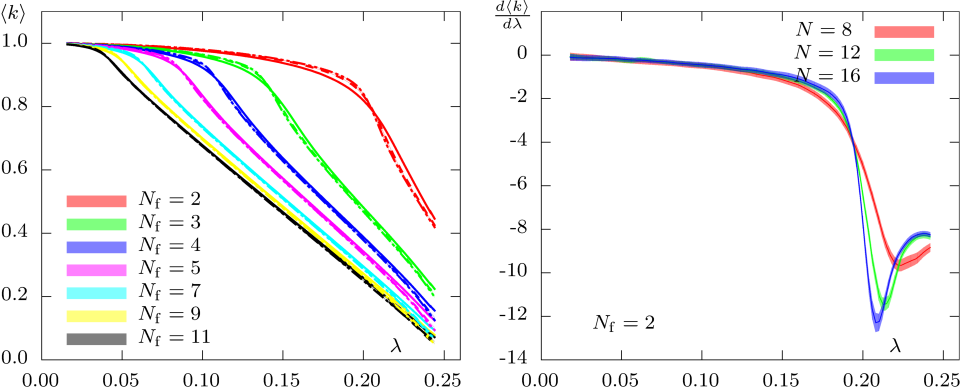}\hskip5mm
 	\caption{Left: average lattice filling factor (\ref{tm23}) on
 	three lattices for every flavor between $2$ and $11$. Right: 
 	With increasing lattice size a singularity of
  the susceptibility (\ref{tm25}) builds up, here for $2$ irreducible flavors.  Figures taken from \cite{WSW17}.\label{fig:wipf_k}}
 \end{figure}

An interesting observable in this context is the lattice filling factor
 \begin{equation}
 k=\frac{1}{2V\Nf}\sum_{i=1}^2\sum_{a,b=1}^{\Nf}\sum_{x=1}^V k_{xi}^{ab}
 \in [0,1]\,,\label{tm21}
 \end{equation}
which relates to the number of states
occupied by the interaction on a lattice site -- 
a natural variable in the dual formulation. It
counts how many fermions take part in the 
non-trivial interaction. In the weak coupling
limit $\lambda\to\infty$
the filling factor vanishes and in the strong coupling limit
$\lambda\to 0$ it is one. It cannot
exceed the value $k=1$ because of Pauli-blocking on every
site. On the left in Figure \ref{fig:wipf_k} we depicted
the average  lattice filling factor, given by a $4$-Fermi
correlation function
\begin{equation}
\langle k\rangle=
-\frac{1}{2\Nf}
\frac{\lambda}{V}\frac{\dot Z(\lambda)}{Z(\lambda)}+c
=\frac{1}{4\Nf\lambda}\langle j^\mu j_\mu\rangle+c\,,\label{tm23}
\end{equation}
for Thirring models with flavor number between $2$ and $11$ 
and lattice sizes $N=8,12$ and $16$. We see clearly that at strong couplings
$\lambda\lessapprox \lambda^*$ the systems are in 
a lattice-artifact phase dominated by Pauli blocking.
We also see that the average filling factor does not suffer
much from finite size corrections, in particular in the strong coupling
region.

The susceptibility associated with $k$ is given by a $8$-Fermi 
correlation function,
\begin{equation}
\partial_\lambda \langle k\rangle  = 
-\frac{1}{16\Nf \lambda^3}\sum_x \big\langle (j^\mu j_\mu)(x)
(j^\mu j_\mu)(0)\big\rangle_c
+\frac{\mathrm{c}-\langle k\rangle}{\lambda}\,,\label{tm25}
\end{equation}
and its dependence on $\lambda$ is seen in 
the right panel of Figure~\ref{fig:wipf_k}
for the model with  $\Nf=2$ on three different lattice
sizes. The dip
of the susceptibility increases with the lattice size
and this points to a transition from a strong-coupling 
lattice artifact phase to a phase which connects to continuum 
physics. 

\subsection{Spontaneous symmetry breaking of parity for odd 
	\texorpdfstring{$\Nf\leq \Nfc$}{textmuu}}\label{subsec2}
\begin{figure}[h]
 	\centering
	\includegraphics[scale=0.95]{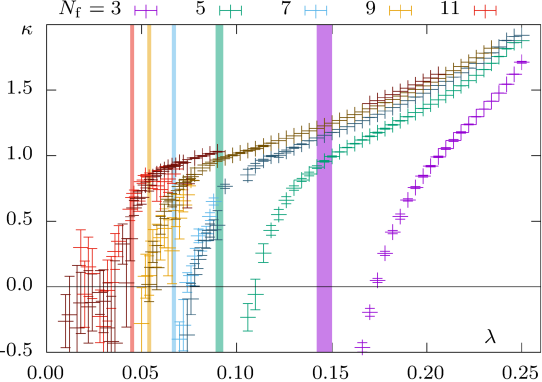}\hskip5mm
	\caption{
		The curvature $\kappa$ of the effective potentials
		at the origin is shown for odd $\Nf$. 
		The curvature is taken in the channel with
		$n=0$, since in the other channels the curvature
		is larger. Figure taken from \cite{WSW17}.\label{fig:curvodd}}
\end{figure}

For odd $\Nf$ the approximate critical coupling $\lambda^*$
on lattices with $N=8,12$ and $16$ are listed in Table~\ref{tab:critcouplings}.
\begin{table}[h!]\centering
	\caption{The critical couplings $\lambda^*$ separating the
		artifact phase and physical phase for three lattice sizes
		and the Thirring lattice models with an odd number of flavors.}\label{tab:critcouplings}
			\begin{tabular}{m{2.2cm}m{2.8cm}m{1.9cm}m{1.9cm}m{1.9cm}m{1.9cm}}\toprule
		$N$&$\Nf=3$&5&7&9&11\\ \midrule
		$8$&$\lambda^*=0.158(4)$&0.098(2)&0.073(2)&0.058(2)&0.048(2)\\
		$12$&$\lambda^*=0.149(4)$&0.094(3)&0.068(2)&0.054(2)&0.046(2)\\
		$16$&$\lambda^*=0.146(4)$&0.091(2)&0.068(2)&0.054(2)&0.046(2)\\
	\bottomrule
	\end{tabular}
\end{table}
A critical coupling $\lambda_c$ of a transition,
where a condensate forms, must be larger than $\lambda^*$ 
to stay away from the artifact phase. We have seen earlier 
that if a condensate forms, then it is the parity-breaking
condensate with $n=\Nf$. This means
that for odd $\Nf$ the system is either in the symmetric phase or
develops a condensate which does not break chiral symmetry
but only parity. Figure~\ref{fig:curvodd} shows the curvature $\kappa$
of the effective potential in the direction with $n=\Nf$ as
function of the inverse coupling. The bars show the estimated 
infinite volume extrapolation of $\lambda^*$, such that only 
values to the right to these bars are in the phase which
connects to continuum physics. For example, for $\Nf=3$
the effective potential becomes unstable against
condensation of $\langle\bar{\psi}_a\psi_b\rangle\propto\delta_{ab}$
at the inverse coupling $\lambda_c\approx0.172$, well outside of the lattice artifact phase
at strong couplings $\lambda\lessapprox\lambda^*\approx 0.145$.

We observe a similar instability against condensation for
all odd flavor numbers between $1\leq \Nf\leq 9$. Actually we 
are not
sure that parity is broken for $\Nf=9$, since the critical couplings
$\lambda^*=0.51$ and $\lambda_c=0.53$, measured on a lattice
with $N=20$, are almost identical. But our simulations clearly
reveal that there is no broken phase for $\Nf\geq 11$ and there
is a broken phase for $\Nf\leq 7$. The critical flavor 
number $\Nfc\approx 9$ for two-component flavors corresponds to a critical flavor 
number $\Nrc\approx 4.5$ for four-component flavors  and thus 
is in good agreement with some of the earlier results in
Table~\ref{tab:ncrit_past}.

\subsection{Special case: \texorpdfstring{$\Nf=1$}{}}
While the models  with even $\Nf$ have no sign problem,
we found numerically that a possible sign 
problem for odd $\Nf$ is extremely mild 
(not observed in our ensembles) for all $\Nf\geq 3$. Only
the irreducible single-flavor Thirring model has a 
severe sign problem that hinders a direct simulation.

In order to still perform simulations, we formulated a
fermion-bag-like algorithm: Starting from the bosonized 
Fierz-transformed formulation (\ref{symm55}),
one  expands the Boltzmann factor in powers of the Yukawa
interaction
\begin{align}
Z=\int\!\cD\sigma\cD\psi\cD\bar{\psi}\;
e^{-S_0[\psi]-\lambda\sum\sigma_x^2}\,\prod_{x,i}\sum_{k_{xi}=0}^{1}\frac{\left(-\sigma_{x}\bar{\psi}_{x}^{i}\psi_{x}^{i}\right)^{k_{xi}}}{k_{xi}!}\,,
\end{align}
where $S_0$ is the lattice action of free SLAC fermions and
for simplicity we already set $m=0$.
The coefficient $\lambda=1/6g^2$ 
of $\sigma^2$ is different from the coefficient $\lambda$ of $v_\mu v^\mu$
in subsections \ref{subsec1} and \ref{subsec2}. The product 
and sum combine into one large sum over sets of occupation 
number variables $k_{xi}$ which only assume the values 
$0$ and $1$ due to the Grassmann nature of $\psi$ and $\psib$.
Now, one can see that whenever $k_{xi}=1$ the Berezin integrals
over $\psi_{x}^{i}$ and $\bar{\psi}_{x}^{i}$ are saturated by the
Yukawa term. The kinetic term must, hence, contribute
trivially which is equivalent to removing the rows and columns $(x,i)$
from $\fdi$. The remaining fermionic and bosonic integrals decouple
leaving a fermion determinant and a bosonic weight factor
\begin{equation}
Z=\sum_{\{k_{x}\}}\left(-1\right)^{kV}w(k)\det\I\fdi[\{k_{x}\}]\,,
\end{equation}
where $\fdi[\{k_{x}\}]$ denotes the reduced fermion operator,
the lattice filling factor $k$ has been defined in (\ref{tm21}) and
\begin{equation}
w(k)=\left(\sqrt{\frac{\lambda}{\pi}}\,\int\!\mathrm{d}\sigma\; \sigma^{2}e^{-\lambda\sigma^{2}}\right)^{kV}.
\end{equation}
One should note that the bosonic integral over $\sigma_{x}$ vanishes
whenever $k_{x1}\neq k_{x2}$. Thus, we used the notation $k_{x}\equiv k_{x1}=k_{x2}$.
This result is similar to the fermion bag algorithm \cite{Chandrasekharan10} with the notable
difference that we do not find "bags" in the sense of connected
clusters due to the non-local nature of chiral SLAC 
fermions.\footnote{A much more detailed derivation of these 
formulae can be found in \cite{lenz18}.}

In order to generate the Markov chain, we use simple Metropolis updates starting from the completely filled lattice with
$\det\I\fdi[\{1,\dots,1\}]=1$. However, we found that flipping a 
single $k_{x}$ often yields configurations of vanishing weight
such that we also propose simultaneous flips of two $k_{x}$.
To speed up the updates of the fermion determinant, we employ a
combination of matrix determinant lemmas and
the Sherman-Morrison-Woodbury formula.

While we can prove that $\det\I\fdi[\{ k_{x}\}]$ is real and positive \cite{lenz18}, the sign factor $\left(-1\right)^{kV}$ 
remains problematic. 
But we found numerically that
\begin{equation}
\left\langle k \right\rangle = \frac{1}{Z}\sum_{\{k_{x}\}}k\left(-1\right)^{kV}w(k)\det\I\fdi[\{k_{x}\}]
\end{equation}
does not suffer the usual increase in variance 
and instead yields a reliable signal. 
This expectation value enters the expression for the
local effective potential $V_{\mathrm{loc}}$. The latter
characterized the distribution function $\exp(-V_\mathrm{loc})$
of $\sigma_x$ and yields the partition function,
\begin{align}
Z=\int\!\mathrm{d}\sigma_{x}\; e^{-V_{\mathrm{loc}}(\sigma_{x})}
\end{align}
for an arbitrary $x$. Due to translational symmetry we assume that
the distribution of $\sigma_{x}$ is independent of $x$.
The potential has the form
\begin{equation}
V_{\mathrm{loc}}(\sigma_{x})=\lambda\sigma_{x}^{2}+ \ln\left(a_{0}+a_{2}\sigma_{x}^{2}\right),
\end{equation}
wherein the coefficients are related in a straightforward
way to expectation values of moments of $k$.
One should stress that $V_{\mathrm{loc}}$ is not quite the standard 
constraint effective potential. But as it describes the statistics of the local
order parameter, we expect it to accurately describe the physics of the
system.

Using the (positive) position of the minimum of the local
effective potential as an order parameter, a detailed study
of the system was performed in \cite{lenz18}. Here, we collect
the main results: The critical coupling was found to be 
\begin{equation}
\label{lambdacnf1}
\lambda_c = \begin{cases}
0.3804(3)&\text{from the condensate},\\
0.3813(3)&\text{from the susceptibility}.
\end{cases}
\end{equation}
The condensate and susceptibility critical exponents are
given as
\begin{equation}
\label{critexpnf1}
\beta = 0.406(8),\quad \gamma=1.1(3).
\end{equation}
The transition to the lattice artifact phase was also analyzed.
It occurred at
\begin{equation}
\lambda^{*} = 0.32838(9).
\end{equation}
There, the data was consistent with a second order phase
transition which is different from the behavior of even flavors.
However, one should note that its equivalence to the single-flavor
GN model might render the $\Nf=1$ Thirring model a special case.

\subsection{No spontaneous symmetry breaking for any even 
	\texorpdfstring{$\Nf$}{} or \texorpdfstring{$\Nr\geq 1$}{}}
\begin{figure}[h]
 		\includegraphics[width=0.48\linewidth]{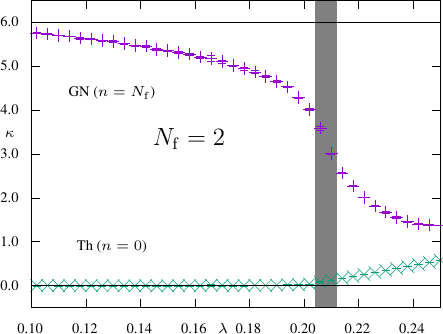}
 	\hfill
 		\includegraphics[width=0.49\linewidth]{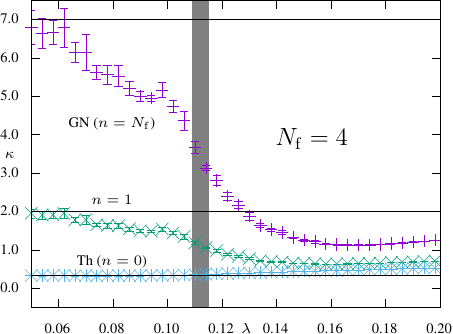}
	\caption{The curvatures $\kappa$ of the effective potentials
		in the channels labelled by the integer $n$ defined in 
		(\ref{tm13}). Left: Thirring model with $\Nf=2$.
		Right: Thirring model with $\Nf=4$.
		The horizontal lines are the results of the strong coupling limits.\label{fig:curveven}}
\end{figure}

For Thirring models with even $\Nf$ -- these are equivalent
to reducible models with $\Nr=\ha\Nf$ -- the situation is
quite different. Figure~\ref{fig:curveven} shows the curvature $\kappa$
of the effective potential $V$ at the origin in the different
channels for $2$ and $4$ irreducible flavors. We see
that for all $\lambda\geq \lambda^*$ the curvatures are
positive such that no condensate can form in the phase which
connects to continuum physics. This striking result
will be further substantiated in the following sections.

\section{Banks-Casher relations}
\begin{figure}[h]
    \includegraphics[width=0.48\linewidth]{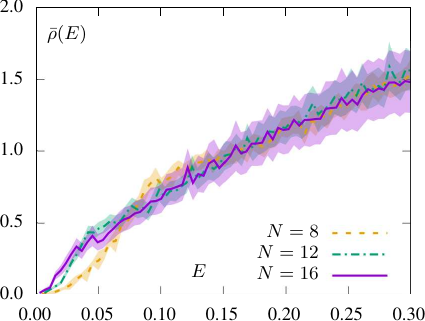}\hfill
    \includegraphics[width=0.48\linewidth]{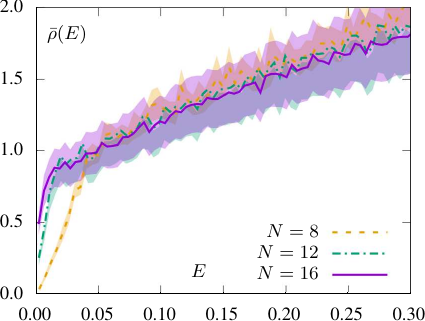}
\caption{The average spectral density $\bar{\rho}$ of the Dirac operator for $1$ reducible flavor (left panel) and for $0.8$ reducible flavors (right panel) \cite{LWW19}.\label{fig:wipf_eigen}
} 
\end{figure}

The effective action of the massive reducible Thirring models with
$\Nf=2\Nr$ reads
	\begin{equation}
	S_\mathrm{eff}=\frac{1}{2g^2}\int\D^3x\, v^\mu v_\mu-\ln 
	\det\big(m^2-\cD_0^2\big)\,,\label{bc1}
	\end{equation}
where $\cD_0$ has been introduced below (\ref{cfn1}). It
is used to investigate the parity-invariant
condensate of the reducible models in the channel with $n=0$, in which the symmetry
U$(2\Nr)$ breaks to $\mathrm{U}(\Nr)\times \mathrm{U}(\Nr)$
and in which the effective potential has minimal curvature.
The parity-even condensate is
	\begin{align}
	\Sigma&=
	\frac{1}{V}\frac{1}{Z}\int\!\mathcal{D}v^{\mu}\;
	\tr\left(\frac{m}{m^2-\cD_0^2}\right)e^{-\Nf S_\mathrm{eff}(v)}\nonumber\\
	&=\frac {2m}{V}\int_0^\infty
	\frac{\D E}{m^2+E^2}\,\bar\rho(E)\label{bc3}
	\end{align}
and it contains the average spectral density $\bar{\rho}$, defined by
\begin{equation}
	\bar\rho(E)=\frac{1}{Z}
	\int\! \mathcal{D}v^\mu\, e^{-\Nf S_\mathrm{eff}(v)}\rho_v(E)\,.\label{bc5a}
\end{equation}
The spectral density of the Dirac operator $\I\cD_0$ for a fixed $v_\mu$ is defined by
\begin{equation}
	\tr\, f(\I\cD_0)=\int_{-\infty}^\infty\! \D E\,f(E)
	\rho_v(E)\,.\label{bc5b}
\end{equation}
The relation (\ref{bc3}) implies that the condensate $\Sigma$
vanishes for $m\to 0$ in case the integral over $E$ is finite.
A condensate forms for small $m$ if the integral
is proportional to $1/m$. This happens when for a typical
configuration there is an abundance of low-lying eigenvalues 
of $-\cD_0^2$. This is the content of the celebrated Banks-Casher
relation which states that in the infinite volume limit
the condensate is proportional to $\bar{\rho}(0)$. 
Figure \ref{fig:wipf_eigen}	shows the average spectral densities 
for $\Nr=1$ (left panel) and for $\Nr=0.8$ (right panel).
Whereas small eigenvalues accumulate for the smaller flavor number
this does not happen for one reducible flavor.
This clearly indicates that there is no spontaneous
breaking of the chiral symmetry in all reducible Thirring models
with $\Nr\geq 1$, in contrast to many previous studies.
It also shows that there is a condensation of small eigenvalues
for $\Nr=0.8$.

The next step beyond using the Banks–Casher relation could be to exploit the (for
non-Abelian gauge theories) well-established link between chiral properties and chiral
random matrix theory (cRMT) which describes universal statistical properties of low-lying
eigenvalues of the Dirac operator \cite{VW00}. 
The constraints imposed by chiral symmetry and its
spontaneous breaking determine the structure of low-energy effective partition function.
The RMT allows for a determination of the chiral condensate from simulations at finite
volume (and nonzero lattice constant a). For SLAC fermions, the lattice Dirac operator
for SLAC fermions is hermitean and in a possible broken phase the statistical properties
of the low-lying Dirac eigenvalues would be determined by the chiral condensate. In
future studies it should be interesting to compare the distribution of individual eigenvalues 
or the spacing between them to RMT prediction to arrive at more stable infinite volume
extrapolations—both for the breaking of parity in the irreducible models and the breaking
of ’chirality’ in the reducible ones.

\section{Spectrum of low-lying states}
If the $\text{U}(2\Nr)$-symmetry were spontaneously broken,
the particle spectrum would reveal the existence of
Goldstone modes. In formulations with \emph{domain wall fermions}
the analysis of the spectrum indicates
SSB for $\Nr=1$. This is in conflict with the results
obtained with the effective potential and spectral density
for \emph{chiral SLAC fermions}, which clearly show that $\Nrc<1$.

To clarify the situations for $\Nr\approx 1$ 
we study the spectrum of light mesons.
The four interpolating operators are
\begin{equation}
\cO_a(x)=\bar\psi(x)(\sigma_a\otimes\sigma_0)\psi(x),\quad
\cO_a(t)=\sum_{\vx}\cO_a(t,\vx),\quad 0\leq a\leq 3\,,
\end{equation}
where $\sigma_0=\id_2$ and $\sigma_1,\sigma_2,\sigma_3$ are the
Pauli-matrices and we use the reducible formulation (\ref{cfn1})
of the Thirring models. Then the expectation values of
$\cO_3$ and $\cO_0$ are identified as chiral and
parity condensates. The latter vanishes in the parity-invariant
reducible models. The correlation matrix of the
interpolating operators is diagonal, $C_{ab}(t)=\langle \cO_a(t)\cO_b(t)\rangle_c=C_a(t)\delta_{ab}$, and is used to
extract the masses of the light mesons. One can show
that the $C_1$ and $C_2$ are equal. If the symmetry U$(2)$ 
is not broken, then we should see a singlet and triplet of U$(2)$
in the spectrum. The correlation
functions $C_1,C_2,C_3$ belong to this triplet. On the other hand, 
if U$(2)$ is spontaneously broken to U$(1)\otimes$ U$(1)$, then we
should detect two Goldstone bosons, related to
the interpolating operators $\cO_1$ and $\cO_2$.

\begin{figure}[h]
		\includegraphics[width=0.48\linewidth]{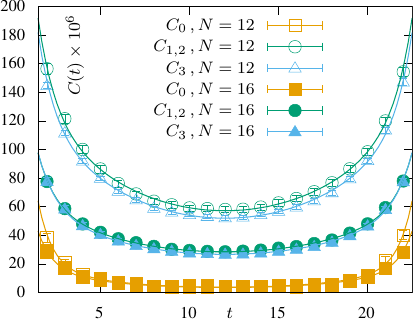}\hfill
		\includegraphics[width=0.48\linewidth]{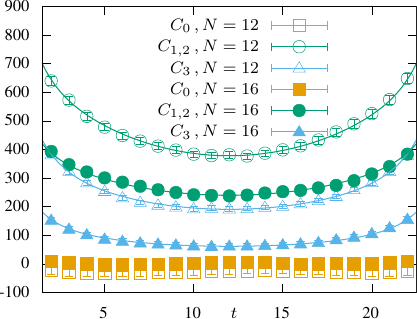}
\caption{The correlation functions $C_0,C_1,C_2$ and $C_3$
	of the interpolating operators for the low-lying mesons
	on lattice of various sizes. On the left for $\Nr=1$ 
	and on the right for $\Nr=0.8$ \cite{LWW19}.\label{fig:spectrum}}
\end{figure}
\unskip

From the correlation functions on a relatively
small lattice with $N=12$ and a larger lattice with $N=16$
we extracted the masses given in Table~\ref{tab:masses}.
We see that for $\Nr=1.0$ the masses extracted from $C_{1,2}$ and $C_3$
are comparable. They belong to the expected triplet in the
symmetric phase. On the other hand, for $\Nf=0.8$ we
do not see such a triplet. Only $m_1$ and $m_2$
are degenerate and they are identified as Goldstone particles.
In the broken phase the correlation function
$C_0$ falls off rapidly and we cannot extract a 
reliable value for the mass
on the relatively small lattices considered.
Hence this mass is missing in the broken phase in the Table.
As earlier we conclude that the critical flavor number $\Nrc$
is smaller than $1$. But we also conclude that it is larger
than $0.8$. 

The next step beyond using the Banks–Casher relation could be to exploit the 
(for non-Abelian gauge theories) well-established link between chiral 
properties and chiral random matrix theory (cRMT) which describes universal statistical properties of low-lying eigenvalues of the Dirac operator 
\cite{VW00}. The constraints imposed by chiral symmetry and its
spontaneous breaking determine the structure of low-energy effective 
partition function.
The RMT allows for a determination of the chiral condensate from simulations 
at finite volume (and nonzero lattice constant a). For SLAC fermions, 
the lattice Dirac operator
for SLAC fermions is hermitean and in a possible broken phase the statistical
properties of the low-lying Dirac eigenvalues 
would be determined by the chiral condensate. In
future studies it should be interesting to compare the distribution 
of individual eigenvalues or the spacing between them to RMT prediction 
to arrive at more stable infinite volume
extrapolations—both for the breaking of parity in the irreducible models and the breaking of ’chirality’ in the reducible ones.

\begin{table}[h] 
	\centering
	\caption{Masses of the lightest mesons on two lattices for
		$\Nr=1.0$ and $0.8$ reducible flavors.\label{tab:masses}}
	\begin{tabular}{m{2.5cm}m{2.2cm}m{2.2cm}m{2.2cm}m{2.5cm}}
		\toprule
		$C$ & $m(12)$ & $m(16)$ & $\Nr$&Symm.\\
		\midrule
		$C_0$ & $0.21(2)$ & $0.21(2)$ & $1.0$&\\
	$C_{1,2}$ & $0.134(3)$ & $0.128(2)$ & $1.0$&
	U(2)\\
	$C_3$ & $0.138(2)$ & $0.131(2)$ & $1.0$&\\
	\midrule
	$C_{1,2}$ & $0.103(2)$ & $0.095(3)$ & $0.8$&
	U(1)$\times$ U(1)\\
	$C_3$ & $0.109(4)$ & $0.127(7)$ & $0.8$&\\
	\bottomrule
	\end{tabular}
\end{table}

\section{Estimating the critical flavor number \texorpdfstring{$\Nrc$}{}}
\begin{figure}[h]
			\includegraphics[width=0.48\linewidth]{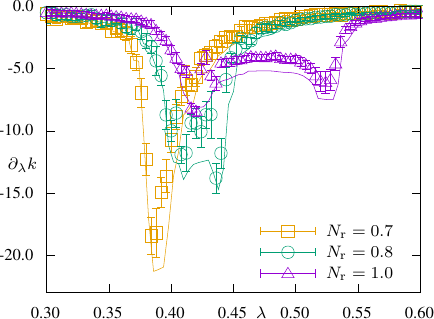}
		\hfill
			\includegraphics[width=0.48\linewidth]{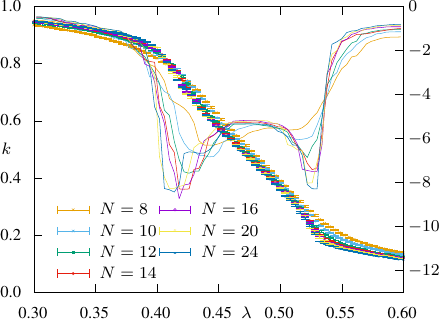}
	\caption{The susceptibility (\ref{tm25}) for Thirring models with
	$\Nr=0.7,0.8$ and $1.0$ reducible flavors (left panel) and 
the volume-dependence of the the average filling factor (\ref{tm23}))
and corresponding susceptibility for $\Nr=1$ (right panel).\label{fig:dips}}
\end{figure}

To extract a reliable estimate for $\Nrc$
we performed a detailed finite size analysis of the susceptibility,
$\partial_\lambda \langle k\rangle$ in (\ref{tm25}) on a
grid of $\Nr$-values between $0.5$ and $1.0$.
The simulation results reveal two dips of the susceptibility
for all $\Nr\gtrapprox  0.78$. Three examples are depicted in
Figure~\ref{fig:dips}, left panel. The dips become more
pronounced with increasing system size, as seen in the same
Figure on the right.
The dip at stronger coupling belongs to the transition into the
lattice artifact phase discussed above. More interestingly, we 
see a second dip at weaker coupling (larger $\lambda$). Most likely
it points to a phase transition without order parameter.

\begin{figure}[h]
  \centering
	\includegraphics[scale=1.1]{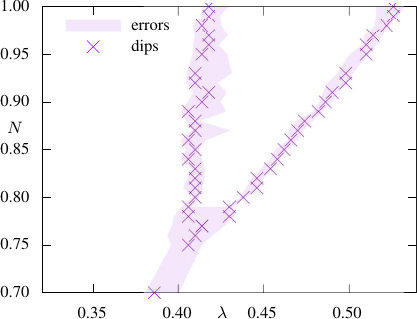}
	\caption{The dip positions of the susceptibility $\partial_\lambda k$.\label{fig:peaks}}
\end{figure}

The positions of the susceptibility dips on a fine
grid in the $(\lambda,\Nr)$-plane near $\Nr\approx 1$
are calculated with an expensive scan of the
susceptibility as function of the inverse coupling
$\lambda$.
The resulting phase transition lines of the (probably first order) ubiquitous 
lattice artifact transitions and of the (probably smooth) new transitions for
all $\Nr\gtrapprox 0.78(4)$ are depicted in Figure~\ref{fig:peaks}.
If this interpretation is correct -- further studies are
needed to answer this question -- 
then one could construct a continuum limit at a
transition without order parameter. Since there is no transition with
order parameter for all reducible lattice Thirring models
with $\Nr\geq 1$, 
such an unusual transition is needed to relate the lattice
models to continuum physics. Transitions without local
order parameter have been reported previously in strongly
coupled Fermi systems \cite{CS17}.
For unrealistically small $\Nr\lessapprox 0.76$ there is one phase 
transition with order parameter. 
Actually, in \cite{LWW19} the maximum of the parity-invariant 
condensate has been measured and the results show clearly
that a condensate forms at these small $\Nr$.

\section{Summary}
In this paper, we reviewed our current knowledge about spontaneous
symmetry breaking in 1+2D Thirring models.
For odd irreducible flavor numbers, these models spontaneously 
break parity symmetry below
\begin{equation}
	\Nf^{\mathrm{crit}}\approx 9.
\end{equation}
For $\Nf=1$, previously unpublished values from \cite{lenz18} for the critical couplings
and critical exponents where given in (\ref{lambdacnf1}) and (\ref{critexpnf1}).

We have collected strong evidence that the critical
flavor number of the reducible Thirring models is 
below $\Nr=1$. We calculated the spectral density, the spectrum 
of scalar and pseudo-scalar mesons as function of the flavor 
number $\Nr$ between $0.5$ and $1.0$ and the maximum of
the chiral condensate \cite{WSW17,LWW19}.  As a result we find a 
critical flavor number 
\begin{equation}
\Nrc= 0.80(4)\,.
\label{final1}
\end{equation}
In particular, we spotted two Goldstone bosons only for
$\Nr\leq \Nrc$.  Since a non-integer value of
$\Nr$ probably does not describe a local quantum field theory, we conclude that there is no SSB in all reducible Thirring models.
The critical value extracted from a combined analysis of
all available data is a bit higher than the value $0.78$
extracted from the susceptibility alone, but the two values
$0.80$ and $0.78$ are compatible within the quoted statistical 
errors.

Simulations based on DWF with a
large extra dimension spot a second order phase 
transition for one reducible flavor.
From a fit to the equation of state the critical exponents
$\delta=4.17(5)$ and $\eta=0.320(5)$ have been estimated \cite{HMW21}.
Similar values have been extracted by the same authors
in a previous study \cite{HMW20}. With DWF
a bilinear condensate forms at the transition, in contrast to
the results obtained with chiral SLAC fermions. On the other hand,
in simulations with DWF 
no transition without order parameter, as monitored
by the fermionic $8$-point function in 
Figure~\ref{fig:dips}, is reported. With ongoing simulations
this issue will hopefully be settled in the near future.

To correctly interpret the lattice results one must stay away from the
lattice-artifact phase which in the dual formulation is
well understood: as for gauge theories at large
chemical potential there is Pauli-blocking on the lattice sites
if the Thirring coupling exceeds a critical value.
Our results are based on dedicated simulations with chiral
SLAC fermions which respect all global inner symmetries and
discrete space-time symmetries, such that there is no doubt that
the lattice models represent the Thirring models in the continuum.
One of the most pressing problem is the nature of the newly
found (probably second order) phase transition without order parameter.
We hope to report on this issue in a future work.

\section*{Acknowledgments}
We would like to thank the organizers
of the workshop  
"Relativistic Fermions in Flatland: theory and application"
"Quantum Theory and Symmetries",
Simon Hands, Holger Gies, John Gracey and Igor Herbut for
organizing an inspiring online meeting at ETC*-Trento.
We thank our former co-workers Daniel Schmidt and Bjoern Wellegehausen 
who contributed considerably to the results presented in
this publication.

\end{document}